\documentclass[nopreprintline,12pt]{elsarticle}

\usepackage{graphicx}%
\usepackage{multirow}%
\usepackage{amsmath,amssymb,amsfonts}%
\usepackage{amsthm}%
\usepackage{mathrsfs}%
\usepackage[title]{appendix}%
\usepackage{xcolor}%
\usepackage{textcomp}%
\usepackage{manyfoot}%
\usepackage{booktabs}%
\usepackage{algorithm}%
\usepackage{algorithmicx}%
\usepackage{algpseudocode}%
\usepackage{listings}%
\usepackage{orcidlink}

\usepackage{bm}
\usepackage{epsfig}
\usepackage{epstopdf}
\usepackage{subfigure}
\usepackage{rotating}

\usepackage{hyperref}

\renewcommand{\bar}{\overline}
\renewcommand{\rho}{\varrho}
\renewcommand{\phi}{\varphi}
\newcommand{\be}{\begin{equation}}
\newcommand{\ee}{\end{equation}}
\newcommand{\dis}{\displaystyle}

\theoremstyle{thmstyleone}%

\theoremstyle{thmstyletwo}%

\theoremstyle{thmstylethree}%
\begin{document}

\begin{frontmatter}



\title{Stabilising effect of generic anomalous diffusion independent of the Rayleigh number}


\author[1]{Antonio Barletta}

\author[1]{Pedro Vayssi\`ere Brand\~ao}

\author[2]{Florinda Capone}

\author[2]{Roberta De Luca}

\affiliation[1]{organization={Dipartimento di Ingegneria Industriale, Alma Mater Studiorum Universita di Bologna}, 
					addresline={Viale del Risorgimento, 2}, 			
					city={Bologna}, 				
					postcode={40136}, 			
					state={}, 			
					country={Italy}}

\affiliation[2]{organization={Dipartimento di Matematica e Applicazioni 'R.Caccioppoli', Universita degli Studi di Napoli Federico II}, 			
					addresline={Via Cintia, Monte S.Angelo, 26}, 			
					city={Napoli}, 			
					postcode={80126}, 			
					state={}, 			
					country={Italy}}


\begin{abstract}
This work investigates the influence of a generic anomalous diffusion model on mass convection in a fluid-saturated porous medium, focusing on superdiffusive regimes. A mathematical model is developed, and stability analyses - both linear and nonlinear - are performed. Results demonstrate that the specific form of the time function describing anomalous diffusion significantly affects system stability, allowing stability to persist beyond the classical Rayleigh-Bénard neutral threshold. Furthermore, transient perturbation growth is observed under certain conditions, followed by eventual decay. The paper systematically examines various memory functions, including power-law, exponential, and logarithmic forms, highlighting their impact on the dynamics of disturbances. The findings underscore the importance of anomalous diffusion in modulating stability and provide new insights into the transient behaviours induced by non-Fickian mass transport.
\end{abstract}



\begin{keyword}
Solutal convection \sep Anomalous diffusion \sep Stability


\end{keyword}

\end{frontmatter}



\section{Introduction}
Anomalous diffusion, also known as non-Brownian diffusion, refers to the deviation from the classical Brownian motion observed in various physical, biological, and chemical systems. Unlike standard diffusion, where the mean squared displacement (MSD) of particles is directly proportional to time, anomalous diffusion is characterized by a MSD that follows a non-linear relationship with time.
Anomalous diffusion arises from various mechanisms such as heterogeneous media, memory effects, and complex interactions, and it is prevalent in systems like turbulent flows, crowded cellular environments, and disordered materials such as porous media \cite{Metzler,DosSantos}. 
The first studies on anomalous diffusion began in 1926 with the work of Richardson \cite{Richardson}. Richardson showed that the equation proposed by Fick for diffusion is not suitable for describing diffusion in turbulent currents and introduced a non-Fickian diffusion equation where the number of neighbours per unit length varies as a function of the distance between neighbours. The issue of anomalous diffusion was revisited by Scher and Montroll in 1975 \cite{SM} to investigate dispersive transport of a charge carrier in amorphous semiconductors. Since then, numerous studies have been carried out on this topic using different analytical approaches: fractional diffusion; nonlinear diffusion equation and a generalized Langevin equation (see \cite{DosSantos} for more details).
In the context of anomalous diffusion, the assumption of thermodynamic equilibrium at a constant temperature $T$ may not hold. Consequently, the principle of equipartition of energy needs to be reformulated as
$$
\dis\frac{m}{2}\left<\left[\dis\frac{d x(t)}{dt}\right]^2\right>=\dis\frac{\Omega}{2} h(t/\tau),
$$
where $m$ is the mass of molecule dissolved in a fluid subjected to thermal agitation, $\Omega$ is a coefficient with the dimension of energy, $h(t/\tau)$ is a positive-definite dimensionless memory function and $\tau$ is a characteristic time constant.
Introducing $Y(t)=\dis\frac{1}{2}\dis\frac{d}{dt}\langle x^2(t) \rangle$, 
the simplest case involves $Y(t)$ being defined by a power-law of time for sufficiently large times, i.e. $Y(t)=D_{pl} r t^{r-1}$. This implies that the variance in the random position of the molecules is given by
$
\sigma^2_x\simeq 2 D_{pl} t^r
$  and the diffusion  coefficient is 
\be\label{Diff}
D=\dis\frac{1}{2}\dis\frac{d\sigma^2_x}{dt}=D_{pl} r t^{r-1},
\ee
 where the generalized diffusion coefficient $D_{pl}$ has the dimension $\mbox{m}^2$s$^{-r}$ and $r$ is the anomalous diffusion exponent. This exponent indicates the nature of the diffusion: if $0<r<1$, the process is subdiffusive, implying hindered motion; if $r>1$, it is superdiffusive, suggesting enhanced, often directed, motion; if $r=1$ one has standard diffusion. {Henry et al. \cite{henry} cite biological media and porous media as possible environments where subdiffusion of power-law type can be observed, while he cites turbulence in plasmas and transport in polymers as phenomena where superdiffusion of the same type can be observed. Wu and Berland \cite{wu} discussed the physical meaning of time-dependent diffusion coefficient of power-law type demonstrating that such a model satisfies the extended diffusion equation. 

Martin et al. \cite{martin} discussed the anomalous diffusion in bed load transport showing a good agreement between a power-law model for the time-dependent diffusion coefficient and experimental results.  Some authors treat the non-Fickian diffusion by considering a nonlinear viscoelastic constitutive equation for the diffusive flux \cite{camera-roda-A,camera-roda-B}. In their study, anonalous diffusion of power-law type is considered to be a possible non-Fickian behaviour for mass transport in polymers. Other experimental observations regarding anomalous diffusion in different problems can be found in \cite{weeks,vilk,ren,tolic,gal}.}
 
The analysis of the effect of anomalous diffusion on the onset of natural convection, and its comparison with results from standard diffusion, has garnered significant attention from the scientific community due to its wide range of applications. 
In particular, the influence of anomalous diffusion of type (\ref{Diff}) on instability in both porous and clear fluids has been examined in 
 \cite{Barletta,BS,BS2}. 
Precisely, in \cite{Barletta} the analysis of the Rayleigh-B\'enard instability due to the mass diffusion in a fluid-saturated horizontal
porous layer is considered. 
Through linear stability analyis, it has been demonstrated that subdiffusion ($r<1$) causes instability in scenarios where standard diffusion predicts stability, whereas superdiffusion ($r>1$) results in stability for any value of the mass diffusion Rayleigh number $Ra$.
In \cite{BS}, the Rayleigh-B\'enard instability in a binary isothermal mixture is investigated. Linear stability analysis and energy method predict instability in the subdiffusive regime and stability for each value of $Ra$ in the superdiffusive regime. 
In \cite{BS2} the thermal convection with anomalous heat diffusion is analyzed. 
 Through the energy method, it is proved that superdiffusion has a stabilizing effect on thermal conduction solution in the cases of 
 a Darcy porous material incorporating variable gravity or penetrative convection effects, in bidisperse porous material and 
in the case of thermosolutal convection.
\\
\noindent
The power-law dependence on time of the mean squared displacement can be observed in various contexts. However, modern microscopic techniques allow for the identification of different anomalous stochastic processes \cite{Metzler2}.\\
\noindent
The effects of a generic memory function $h(t/\tau)$ (which may not necessarily vary as a power of time) on the stability of fluid motion have yet to be explored. In this work, we aim to investigate the impact of a generic anomalous diffusion law  on mass convection in a fluid saturating a porous medium.
We show that, depending on the time function  describing anomalous diffusion, stability can be maintained even beyond the neutral stability threshold of the Rayleigh-B\'enard problem with standard diffusion. Specifically, there can exist regions of transient growth in perturbations, with their duration and extent determined by the control parameters and the shape of the anomalous diffusion memory function, after which the perturbations decay to zero. Here we focus on generic anomalous diffusion of superdiffusive type.

The paper is organized as follows. Section \ref{sec2} is devoted to the superdiffusion as a mechanism for instability control, by presenting a case study. Section \ref{sec3} introduces the mathematical model describing the flow of a binary mixture saturating a horizontal porous layer with a generic anomalous mass diffusion. The dimensionless perturbation system to the basic state is determined. Section \ref{sec:energy-stab} deals with the nonlinear stability analysis in the energy norm. A sufficient condition guaranteeing the stabilizing effect of generic anomalous diffusion, \emph{for every value of the mass diffusion Rayleigh number}, is determined.
The linear stability analysis is performed in Section \ref{sec5}. Under certain conditions, it is proved that there exists a transient growth of perturbations, before they asymptotically decay to zero.
Different memory functions (power-law, exponential, logarithmic) are considered in Section \ref{sec6}. Some figures with the linear dynamics of disturbances are showed in Section \ref{sec7}.
The paper ends with Section \ref{sec8} containing some concluding remarks.
\section{Superdiffusion as a mechanism for instability control}

\label{sec2}

The apparent stabilising role of superdiffusion on the onset of convection is extensively discussed in the papers \cite{Barletta,BS,BS2}. Linear and nonlinear analyses seem to agree that in an asymptotic sense disturbances are always stable in superdiffusive convection.  We recall that when natural convection takes place, an enhancement of heat transfer is observed, which can be of interest for some applications while undesirable for some others. In this sense, a good comprehension of the role of non-standard diffusion processes on convection can be the key for the development of a framework to control instabilities and consequently the onset of natural convection.

As a case study, one can envisage a salt gradient solar pond (SGSP). It is an artificial pond that has the ability to store solar energy due to its non-convective nature \cite{rghif,nielsen}. Such an ability is a consequence of the presence of an imposed stable vertical gradient of salt concentration (detailed information can be found in \cite{nielsen}). Some efforts were made in the last decades in order to understand better the convection phenomenon in a SGSP \cite{kaffel,suarez}. Karim et al. \cite{karim} studied the linear stability of the gradient zone in a solar pond by considering a model based on clear fluids. In their study, the role of the thermal and concentration gradients ratio is put in evidence. As pointed out in the theoretical work by Nield \cite{nield}, the linear instability threshold of a thermohaline convection in porous media is given by a combination between the dimensionless temperature and concentration gradient. The principle of a SGSP perfomance is the imposition of a stable salt concentration gradient in order to counterbalance an unstable thermal gradient and thus avoid the onset of heat transfer by convection. This means that there exists an upper limit for the thermal gradient that guarantees a pure convective state, whose value depends on the gradient of salt concentration magnitude. Now, if one substitutes the salt with a generic solute whose diffusion is anomalous of superdiffusive type, it would be ideally possible to establish a pure conductive state even in the presence of a very large thermal gradient. It is clear that the operation of a SGSP has other technological limitations besides the presence or not of heat transfer by convection. However, the present exercise is meant to demonstrate the potentiality of a good comprehension of the natural convection phenomena in the presence of anomalous diffusion.

\section{Mathematical model}

\label{sec3}

Let us consider an isothermal, incompressible, Newtonian fluid saturating a horizontal porous layer of depth $d$, subjected only to the gravity force, where a solute is dissolved in. 
The momentum, mass balance and solute transfer equations in a reference frame $\mathcal R=\{O,\mathbf i,\mathbf j,\mathbf k\}$ ($\mathbf k$ pointing vertically upward), are given by 
\be\label{mode}
\begin{cases}
 \dis\frac{\mu}{k}\mathbf v=-\nabla p+\rho_0 g\alpha_c(C-C_0)\mathbf k,\\
 \nabla\cdot\mathbf v=0,\\
 \Phi \dis\frac{\partial C}{\partial t}+\mathbf v\cdot\nabla C=\Phi D_f h(t/\tau)\Delta C,
\end{cases}\ee
where $\mathbf v$ is the seepage velocity, $C$ the solute concentration, $p$ is the local difference between the pressure and the hydrostatic pressure, $t$ is the time, $\mu$ is the dynamic viscosity, $k$ is the permeability, $\rho_0$ is the reference fluid density at the reference concentration $C_0$, $\mathbf g=-g\mathbf k$ is the gravity vector, $\alpha_c$ is the volumetric expansion coefficient, $\Phi$ is the porosity and $D_f$ is the  generalized diffusion coefficient.
To (\ref{mode}), we append the boundary conditions
\be\label{boundT}
\mathbf v\cdot\mathbf n=0,\,\,\mbox{on } z=0,d, \quad C(x,y,0,t)=C_l,\,\, C(x,y,d,t)=C_u,\,\, C_l>C_u.\ee 
Introducing the transformation
\be\label{adim}
\begin{array}{l}
 \mathbf x_{\mathrm{nd}}=\dis\frac{\mathbf x}{d},\,\,
 t_{\mathrm{nd}}=\dis\frac{t}{\tau},\,\, \mathbf v_{\mathrm{nd}}=\dis\frac{\mathbf v \tau}{d},\,\, C_{\mathrm{nd}}=\dis\frac{C-C_l}{C_l-C_u}, \,\, p_{\mathrm{nd}}=\dis\frac{p\tau k}{\mu d^2},
\end{array}\ee
where $\tau= d^2/\Phi D_f$, the dimensionless model is given by (omitting the subscripts)
\be\label{nodim}
\begin{cases}
 \mathbf v=-\nabla p+ Ra C\mathbf k,\\
 \nabla\cdot\mathbf v=0,\\
 \Phi \dis\frac{\partial C}{\partial t}+\mathbf v\cdot\nabla C=h(t) \Delta C,
\end{cases}\ee
where 
\be
Ra=\dis\frac{k\rho_0 \alpha_c  g(C_l-C_u) d}{\Phi \mu D_f }\ee
is the mass diffusion Rayleigh number. 
The boundary conditions (\ref{boundT}) become
\be
\mathbf v\cdot\mathbf k=0,\,\,\mbox{on } z=0,1,\qquad C(x,y,0,t)=0,\quad C(x,y,1,t)=-1.\ee
The basic motion is given by
\be \label{eq:basic-state}
\bar{\mathbf v}=\mathbf 0,\quad \bar C(z)=-z,\quad \bar p=\bar p_0-Ra\dis\frac{z^2}{2}.
\ee
Introducing the perturbation fields 
\be
\mathbf u=\mathbf v-\bar{\mathbf v},\quad \Pi=p-\bar p,\quad \Gamma=C-\bar C,\ee
the non-dimensional perturbation system is given by
\be\label{pertsys}
\begin{cases}
 \mathbf u=-\nabla\Pi+ Ra\Gamma\mathbf k,\\
 \nabla\cdot\mathbf u=0,\\
\Phi \dis\frac{\partial\Gamma}{\partial t}+\mathbf u\cdot\nabla\Gamma=w+h(t)\Delta \Gamma,
\end{cases}\ee
where $\mathbf u=(u,v,w)$,
under the boundary conditions
\be
w=\Gamma=0,\quad \mbox{on } z=0,1.\ee

\section{Stabilizing effect of generic anomalous diffusion}

\label{sec:energy-stab}

Let us assume that $\{u,v,w,\Gamma,\nabla\Pi\}$ are periodic in the horizontal directions of period $2\pi/a_x$ and $2\pi/a_y$ along the $x$ and $y$ axes respectively and denote by $V=\left[0,\dis\frac{2\pi}{a_x}\right]\times\left[0,\dis\frac{2\pi}{a_y}\right]\times[0,1]$ the periodicity cell, by $\langle \cdot,\cdot \rangle$ and $\|\cdot\|$ the $L^2(V)-$scalar product and norm respectively. 
Multiplying (\ref{pertsys})$_1$ by $\mathbf u$, (\ref{pertsys})$_3$ by $\Gamma$, integrating over $V$ and adding the obtained results, one has 
\be
\begin{array}{l}
 \dis\frac{\Phi}{2}\dis\frac{d}{dt}\|\Gamma\|^2=-
 \|\mathbf u\|^2+(Ra+1)\langle \Gamma,w \rangle- h(t)\|\nabla\Gamma\|^2.
\end{array}\ee
In view of the Poincar\`e and Cauchy inequalities, one recovers that
\be\begin{array}{ll}
 \dis\frac{\Phi}{2}\dis\frac{d}{dt}\|\Gamma\|^2
 &\leq -\|\mathbf u\|^2+\dis\frac{(Ra+1)^2}{4}\|\Gamma\|^2+\|w\|^2- h(t)\pi^2\|\Gamma\|^2\\
& \leq \left[\dis\frac{( Ra +1)^2}{4}-h(t)\pi^2\right]\|\Gamma\|^2. 
   \end{array}
\ee
Then, one has 
\be
\dis\frac{d}{dt}\|\Gamma\|^2+\left[\dis\frac{2\pi^2 h(t)}{\Phi}-\dis\frac{(Ra+1)^2}{2\Phi}\right]\|\Gamma\|^2\leq 0,\ee
i.e.
\be
\dis\frac{d}{dt}\left\{\|\Gamma\|^2\exp\left[
\dis\frac{2\pi^2 H(t)}{\Phi}-\dis\frac{(Ra+1)^2}{2\Phi}t\right]\right\}\leq0,\ee
being $H(t)=\dis\int_0^t h(\tau)\,d\tau$.
Hence
\be
\|\Gamma\|^2\leq \|\Gamma(0)\|^2\exp\left\{
\dis\frac{(Ra+1)^2}{2\Phi}t-\dis\frac{2\pi^2 H(t)}{\Phi}\right\}.\ee
Therefore, if 
\be\label{final}
\dis\lim_{t\to\infty} \dis\frac{H(t)}{t}=\infty,\ee
one has that $\|\Gamma\|^2$ tends to zero for every value of $Ra$.\\
\noindent
From $\|\mathbf u\|^2=Ra\langle\Gamma,w\rangle$, by applying the Cauchy inequality, one easily recovers that, when $\|\Gamma\|^2$ tends to zero, then $\|\mathbf u\|^2$ tends to zero too.
\\
\noindent
Summarizing, when the mass diffusion can be expressed as 
an arbitrary positive, integrable function of time, there is stability for every value of the mass diffusion Rayleigh number if  (\ref{final}) holds,
where $H^\prime(t)=h(t)$.

\section{Linear stability analysis}

\label{sec5}

Section \ref{sec:energy-stab} showed that, depending on the characteristics of the function $h(t)$, the nonlinear stability in the energy norm is ensured for every value of $Ra$. In order to understand better the effect of a generic anomalous diffusion on the possible transition to instability, we investigate in this section the linear dynamics of infinitesimal disturbances when superposed to the basic solution given by Eq.~\eqref{eq:basic-state}. 

Getting rid of the pressure field, in view of the periodicity in the horizontal directions and the boundary conditions, one can write
\be
\mathbf u = \mathbf u_t(t)  \sin (n \pi z) e^{i(\alpha x + \beta y)}, \quad \Gamma = \Gamma_t(t) \sin (n \pi z) e^{i(\alpha x + \beta y)},
\ee
with $\mathbf u_t(t)=(u_t(t),v_t(t),w_t(t))$.
Substituting in the linearized version of (\ref{pertsys}),
algebraic manipulation allows one to arrive at the system of equations for $w_t(t)$ and $\Gamma_t(t)$

\be \label{eq:lin_wt}
w_t = \frac{Ra (\alpha^2 + \beta^2)}{n^2 \pi^2 + \alpha^2 + \beta^2} \Gamma_t,
\ee
\be \label{eq:lin_gammat}
\Phi \frac{\partial \Gamma_t}{\partial t} + (n^2 \pi^2 + \alpha^2 + \beta^2) h(t) \Gamma_t = w_t.
\ee

After substituting Eq.~\eqref{eq:lin_wt} in Eq.~\eqref{eq:lin_gammat}, it yields a single equation for $\Gamma_t$

\be \label{eq:lin_gammat_2}
\frac{\partial \Gamma_t}{\partial t} + \mathcal{C}_1 h(t)\Gamma_t - \mathcal{C}_2 \Gamma_t=0,
\ee
where, $\mathcal{C}_1$ and $\mathcal{C}_2$ are positive coefficients given by 

\be\label{c1}
\mathcal{C}_1 = \frac{1}{\Phi}(n^2 \pi^2 + \alpha^2 + \beta^2),
\ee
\be\label{c2}
\mathcal{C}_2 = \frac{1}{\Phi} \frac{Ra(\alpha^2+\beta^2)}{n^2\pi^2+\alpha^2+\beta^2}.
\ee

The solution of Eq.\eqref{eq:lin_gammat_2} depends on the function $h(t)$, and it is given by

\be
\Gamma_t = \Gamma_0 \, e^{\mathcal{I}(t)},
\ee
where $\Gamma_0 = \Gamma(0)$ and
\be
\mathcal{I}(t) = \int_0^t\left[ \mathcal{C}_2 - \mathcal{C}_1 h(\xi) \right]\mathrm{d}\xi.
\ee

\subsection{Asymptotic behaviour}

The behaviour of $\Gamma_t$ in the asymptotic limit of $t \to \infty$ can be deduced from the behaviour of the function $\mathcal{I}(t)$, which is given by

\be
\mathcal{I}(t)=\mathcal{C}_2 t - \mathcal{C}_1 H(t),
\ee
where $H(t) =\dis\int_0^t h(\tau)\,d\tau$. Considering that $\mathcal{C}_1$ and $\mathcal{C}_2$ are independent of time, and are about the same order of magnitude, we can say that if

\be
\label{eq:lin-limit}
\lim_{t\to\infty} \dis\frac{H(t)}{t}=\infty,
\ee
then 

\be
\lim_{t\to\infty} \Gamma_t=0,
\ee
and the problem is said to be stable from a linear point of view.       \\

\subsection{Transient behaviour}

We have found a sufficient condition to asymptotic stability from a linear point of view, which is given by Eq.~\eqref{eq:lin-limit} and depends on the function $h(t)$. At this point, one could be interested in understanding the short-time behaviour of the disturbances besides their asymptotic behaviour. By looking at the time derivative of $\Gamma_t$ at $t=0$ it is possible to establish if the disturbances have a monotonic or a non-monotonic behaviour in time. In fact, if such a derivative is positive, the behaviour is a non-monotonic one. The time derivative of $\Gamma_t$ at $t=0$ is given by

\be
\left.{\frac{\partial \Gamma_t}{\partial t}} \right|_{t=0} =  \mathcal{C}_2 - \mathcal{C}_1 h(0),
\ee 
which yields a positive derivative for $\mathcal{C}_2 > \mathcal{C}_1 h(0)$.

\subsection{Maximum transient growth}

In the case of non-monotonic behaviour, the maximum transient growth can be determined by evaluating the following equation

\be
\mathcal{C}_2 - \mathcal{C}_1 h(t_{gmax}) = 0,
\ee
where $t_{gmax}$ is the time instant of the maximum transient growth $\Gamma_{tmax}$.

\section{Different memory functions}

\label{sec6}

In this Section, different memory functions will be considered in order to understand the linear dynamics of the disturbances both from an asymptotic and a transient point of view. 

\subsection{Power-law}

\label{sec:power-law}

The memory function can be modelled by a power-law function

\be
\label{eq:power-law-func}
h(t) = r t^{r-1},
\ee
where $r>0$. Two different behaviours are possible. If $r>1$, the process is said to be superdiffusive, while if $r<1$ it is subdiffusive. The case $r=1$ yields the classical diffusion. For $r<1$,
\be
\lim_{t\to\infty} \dis\frac{H(t)}{t}=0,
\ee
meaning that 
\be
\lim_{t\to\infty} \Gamma_t=\infty,
\ee
which implies that for the subdiffusive case disturbances always grow in the asymptotic regime. In this case, we are not interested in the transient behaviour. For $r>1$,
\be
\lim_{t\to\infty} \dis\frac{H(t)}{t}=\infty,
\ee
meaning that 
\be
\lim_{t\to\infty} \Gamma_t=0,
\ee
which implies that the superdiffusive case is always stable from an asymptotic point of view. In this case, it could be interesting to understand the short-time behaviour of the disturbances. For $t=0$, we have $h(0)=0$, and then
\be
\left.{\frac{\partial \Gamma_t}{\partial t}} \right|_{t=0} =  \mathcal{C}_2,
\ee
which is a positive value. Such a result implies that the linear temporal dynamics of the disturbances is a non-monotonic one. The disturbance amplitude grows for small values of $t$ and decays to zero in its asymptotic limit. In this case, the maximum transient growth is observed for

\be
\label{eq:tmax-power-law}
t_{gmax}=\dis\frac{\mathcal{C}_2}{\mathcal{C}_1 r}^{\frac{1}{r-1}}.
\ee

At this instant of time, the disturbance magnitude reaches the value

\be
\label{eq:max-power-law}
\Gamma_{tmax} = \exp \left(\mathcal{C}_2 t_{gmax} - \mathcal{C}_1 t_{gmax}^r \right).
\ee

\subsection{Exponential}

\label{sec:exp}

Another possible way to model the anomalous diffusion is considering a memory function of exponential type, which can be defined by
\be
\label{eq:exp-func}
h(t) = e^{t}.
\ee

In this case, we have
\be
\lim_{t\to\infty} \dis\frac{H(t)}{t}=\infty,
\ee
which implies 
\be
\lim_{t\to\infty} \Gamma_t=0.
\ee

In other words, an exponential memory function yields linearly asymptotically stable disturbances. For $t=0$, we have $h(0)=1$, and then
\be
\left.{\frac{\partial \Gamma_t}{\partial t}} \right|_{t=0} =  \mathcal{C}_2 - \mathcal{C}_1,
\ee
which means that disturbances time derivative at $t=0$ can be either positive or negative depending on the values of the control parameters and the disturbance characteristics. Namely, if $\mathcal{C}_2>\mathcal{C}_1$ disturbances grow at $t=0$, otherwise they decay. If a growth is observed for $t=0$, the temporal behaviour of the disturbances is non-monotonic, since we know that $\lim_{t\to\infty} \Gamma_t=0$. In this case, there is an initial transient growth of the disturbances, followed by an asymptotic decay. The maximum transient growth is achieved here at 

\be
\label{eq:tmax-exp}
t_{gmax} = \log(\mathcal{C}_2/\mathcal{C}_1).
\ee

At this instant of time, the disturbance magnitude reaches the value of 

\be
\label{eq:max-exp}
\Gamma_{tmax} = \left( \frac{\mathcal{C}_2}{\mathcal{C}_1}\right)^{\mathcal{C}_2} \exp\left(\mathcal{C}_1 -\mathcal{C}_2 \right).
\ee

\subsection{Logarithmic}

\label{sec:log}

If a logarithmic behaviour is the appropriate one to model the memory function, it can be defined as

\be
\label{eq:log-func}
h(t) = \log(a+t),
\ee
with $a \geq 1$ being a dimensionless anomalous diffusion parameter for the logarithmic memory function. It is important to remark here that the presence of the parameter $a$ in the definition of the logarithmic memory function has two main reasons: the first one is the necessity of having a positive definite memory function for all values of $t \geq 0$; the second one is the interest in considering a logarithmic function without any loss of generality, if one considers a priori a constant value for $a$, as for instance $a=1$, it would imply a less general model. In this case, we have

\be
\lim_{t\to\infty} \dis\frac{H(t)}{t}=\infty,
\ee
which implies 
\be
\lim_{t\to\infty} \Gamma_t=0.
\ee

For $t=0$, we have that $h(0)=\log(a)$, and consequently

\be
\left.{\frac{\partial \Gamma_t}{\partial t}} \right|_{t=0} = \mathcal{C}_2 - \mathcal{C}_1 \log(a).
\ee

In the case $a=1$, the transient growth is always observed because the disturbance time derivative at $t=0$ is equal do $\mathcal{C}_2$. However, for $a>1$ it depends on the sign of $\mathcal{C}_2 - \mathcal{C}_1 \log(a)$. If $a = \mathrm{e}$ the transient growth is observed just for the case $\mathcal{C}_2 > \mathcal{C}_1$.

This means that the disturbances start to grow at $t=0$ just for some parametric combinations. In the case of transient growth, the maximum growth is achieved for 

\be
\label{eq:tmax-log}
t_{gmax} = e^{\mathcal{C}_2/\mathcal{C}_1} -a,
\ee

At $t=t_{gmax}$ for the present case, the maximum value reached by the disturbances after the transient growth is
\be
\label{eq:max-log}
\Gamma_{tmax} = \exp (a\, \mathcal{C}_1) \exp\left[\left( \exp\left( \frac{\mathcal{C}_2}{\mathcal{C}_1}\right) -a\right) \left(\mathcal{C}_1 + \mathcal{C}_2\right) \right] \exp\left( \frac{\mathcal{C}_2}{\mathcal{C}_1}\right)^{-\mathcal{C}_1 \exp\left( \frac{\mathcal{C}_2}{\mathcal{C}_1}\right)}.
\ee

\section{Discussion of results}

\label{sec7}

In order to analyse the stabilizing effect of different memory functions used to model the anomalous diffusion behaviour, we focused both on the transient and on the asymptotic behaviour of the disturbances. In the case of asymptotic decay of the disturbances, it is imperative to understand their short-time behaviour. In this regime, it is important to know if the disturbances initially start to grow. If so, the time necessary to reach their maximum growth is an interesting additional information. In this section, we compare the behaviour of the different memory functions for different parametric combinations. For the sake of simplicity, without any loss of generality, we consider for all the following results $\Gamma_0 = 1$.

\begin{figure}
\centering
\includegraphics[width=0.45\textwidth]{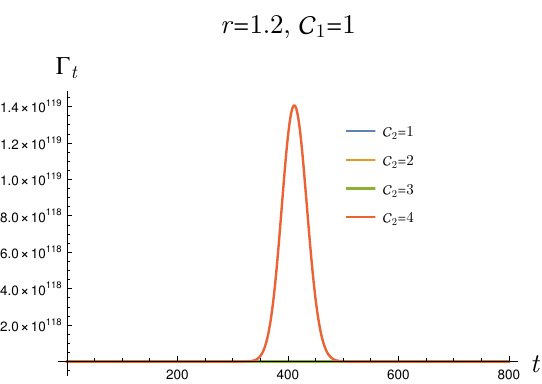}
\includegraphics[width=0.45\textwidth]{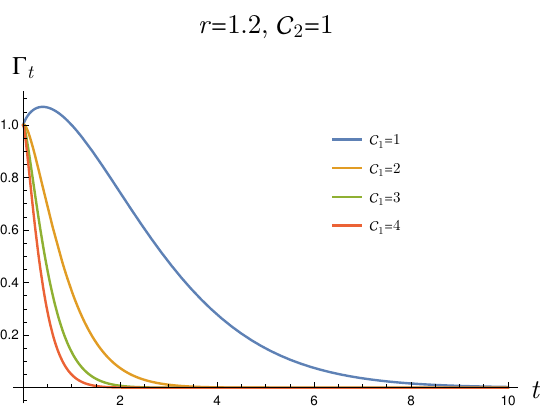}
\caption{\label{fig:power-law-r1p2} Temporal disturbance behaviour for power-law memory function, with $r=1.2$}
\end{figure}

\begin{figure}
\centering
\includegraphics[width=0.45\textwidth]{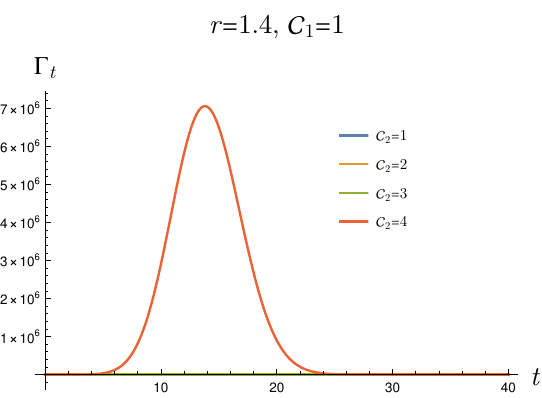}
\includegraphics[width=0.45\textwidth]{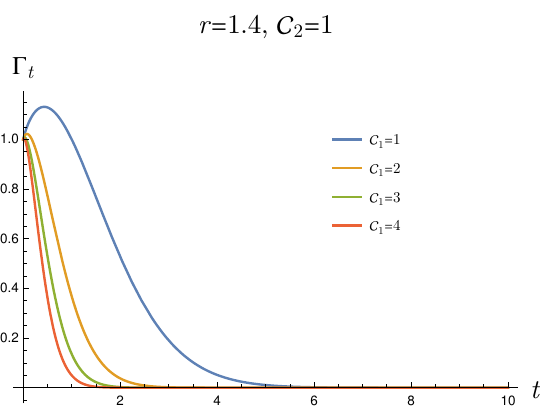}
\caption{\label{fig:power-law-r1p4} Temporal disturbance behaviour for power-law memory function, with $r=1.4$}
\end{figure}

\begin{figure}
\centering
\includegraphics[width=0.45\textwidth]{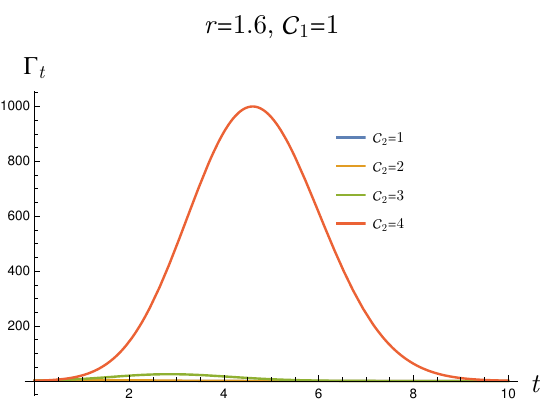}
\includegraphics[width=0.45\textwidth]{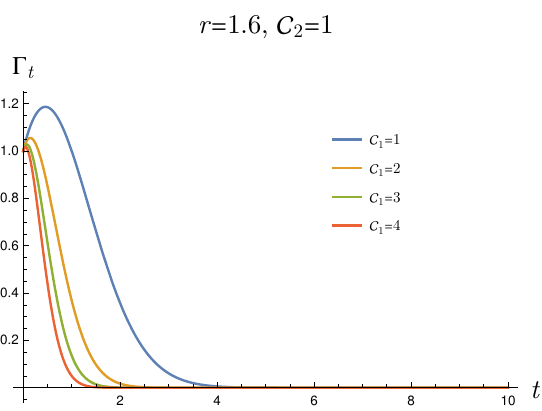}
\caption{\label{fig:power-law-r1p6} Temporal disturbance behaviour for power-law memory function $r=1.6$}
\end{figure}

Figures \ref{fig:power-law-r1p2}, \ref{fig:power-law-r1p4} and \ref{fig:power-law-r1p6} show the temporal behaviour of the disturbances considering the case of a power-law memory function, as described by Eq.\eqref{eq:power-law-func}. All the features related to temporal behaviour of the disturbances discussed in the Section \ref{sec:power-law} are confirmed here. Namely, the asymptotic decay and the initial transient growth of the disturbances, which confirm their non-monotonic behaviour. Another interesting information is the one related to the dimensionless time necessary for the disturbances to reach their maximum value after the transient growth, before the asymptotic decay. By looking at Figures \ref{fig:power-law-r1p2}, \ref{fig:power-law-r1p4} and \ref{fig:power-law-r1p6} it is clear that the time necessary to reach the maximum transient growth is higher for the cases with $\mathcal{C}_2>\mathcal{C}_1$ (left frames). These figures confirm the result given by Eq.\eqref{eq:tmax-power-law}, which states that when $\mathcal{C}_2>\mathcal{C}_1$ the time necessary to reach the maximum growth is higher than in the case $\mathcal{C}_1>\mathcal{C}_2$. In that case, the maximum value reached by the disturbances is higher as well, and it is given by Eq.~\eqref{eq:max-power-law}.

\begin{figure}
\centering
\includegraphics[width=0.45\textwidth]{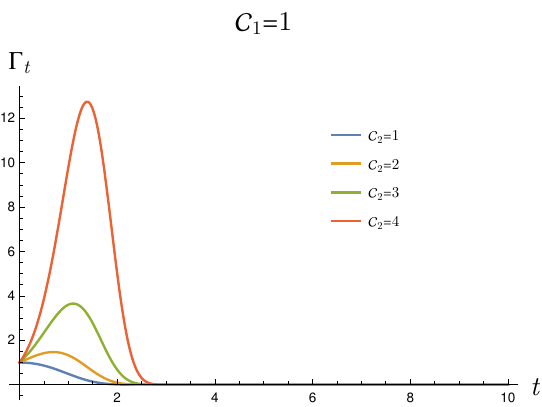}
\includegraphics[width=0.45\textwidth]{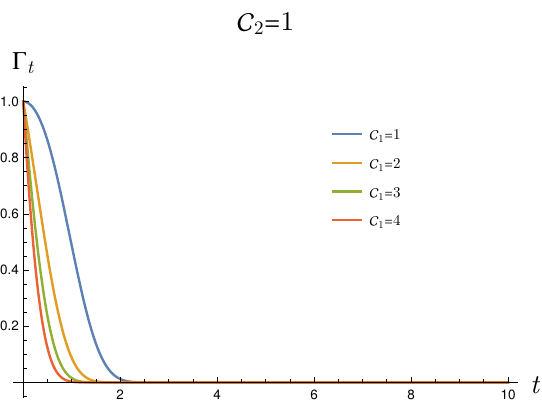}
\caption{\label{fig:exp} Temporal disturbance behaviour for exponential memory function}
\end{figure}

Figure \ref{fig:exp} shows the temporal disturbances dynamics for the case of an exponential memory function, as described by Eq.\eqref{eq:exp-func}. Again, all the features discussed in Section \ref{sec:exp} are confirmed. The  transient growth only exists here for $\mathcal{C}_2>\mathcal{C}_1$. In addition, for higher values of $\mathcal{C}_2$ both the time necessary for the disturbances to reach their maximum values and the maximum values of the magnitude increase.

\begin{figure}
\centering
\includegraphics[width=0.45\textwidth]{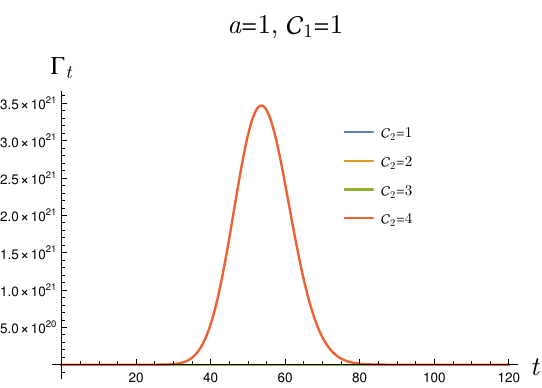}
\includegraphics[width=0.45\textwidth]{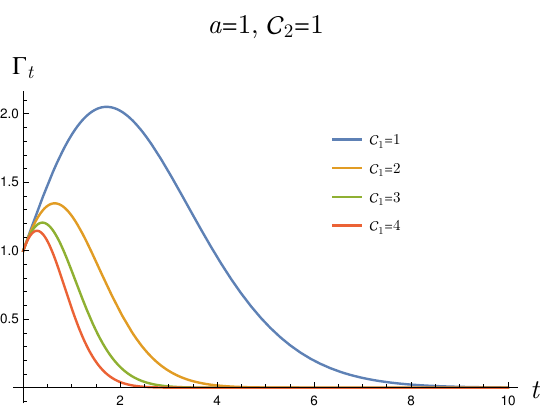}
\caption{\label{fig:log-a1} Temporal disturbance behaviour for logarithmic memory function with $a=1$}
\end{figure}

\begin{figure}
\centering
\includegraphics[width=0.45\textwidth]{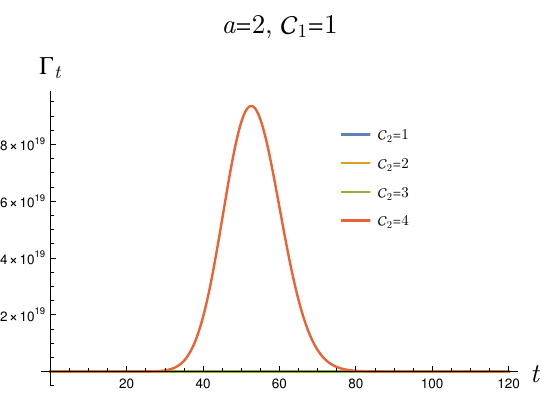}
\includegraphics[width=0.45\textwidth]{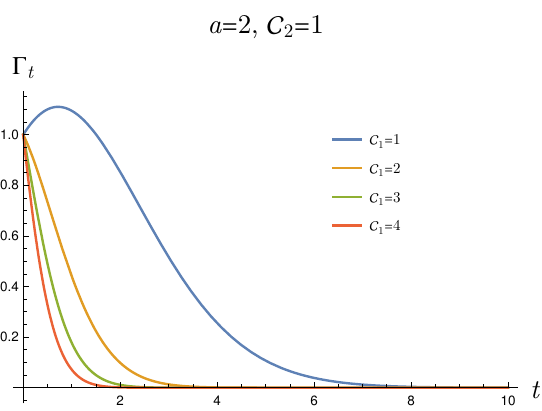}
\caption{\label{fig:log-a2} Temporal disturbance behaviour for logarithmic memory function with $a=2$}
\end{figure}

\begin{figure}
\centering
\includegraphics[width=0.45\textwidth]{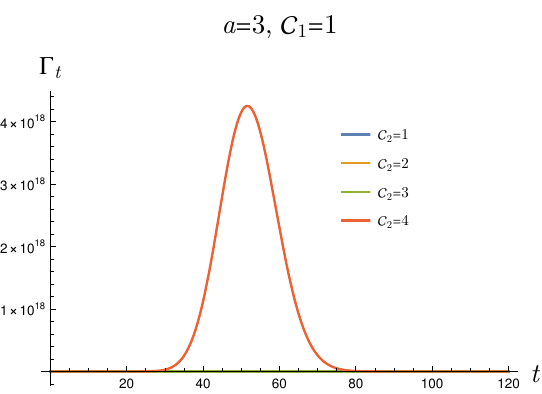}
\includegraphics[width=0.45\textwidth]{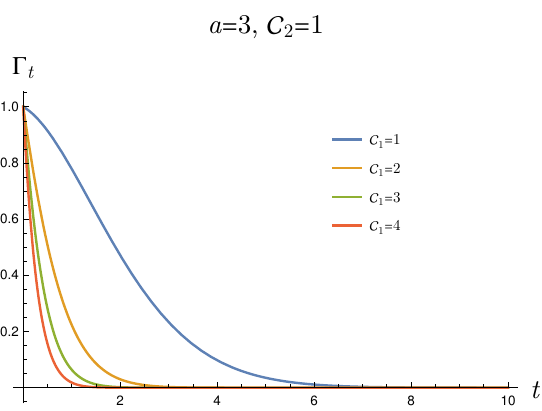}
\caption{\label{fig:log-a3} Temporal disturbance behaviour for logarithmic memory function with $a=3$}
\end{figure}

Figures \ref{fig:log-a1}, \ref{fig:log-a2} and \ref{fig:log-a3} show the linear disturbances dynamics when considering a logarithmic memory function for different values of the parameter $a$. For the logarithmic case, the existence of the transient growth depends on the control and the anomalous diffusion parameters. For $\mathcal{C}_2>\mathcal{C}_1$ both the maximum value of the disturbances magnitude and the time necessary to reach it are higher. In  the case of $\mathcal{C}_2>\mathcal{C}_1$, increasing $\mathcal{C}_1$ yields a smaller $t_{gmax}$ but a larger maximum disturbance magnitude. Increasing the anomalous diffusion parameter $a$ reduces the maximum transient growth as well as the time necessary to reach it. If $a=1$ then the transient growth is always observed. On the other hand, if $a>1$, it is observed just for $\mathcal{C}_2>\mathcal{C}_1 \log(a)$ .

\section{Conclusions}

\label{sec8}

In this paper, we have investigated the influence of anomalous mass diffusion on the onset of convection in a porous medium. We showed that, if different memory functions, that can be used to model the anomalous diffusion, have similar characteristics, some similar conclusions can be made despite their specific form. One of the most important conclusions of the present study, is that for the memory functions tested here, an asymptotic stability is always observed. However, for some prescribed parametric combinations and characteristics of the memory function, a transient growth can take place before the asymptotic decay of the disturbances. 
\\
\noindent
Precisely, when $h(0)=1$ (which is represented here by the exponential memory function), the existence of a transient growth is strictly dependent on the sign of the difference $\mathcal C_1-\mathcal C_2$.
From (\ref{c1})-(\ref{c2}), it easily follows that
$$
\mathcal C_2-\mathcal C_1=\dis\frac{1}{\Phi (a^2+n^2\pi^2)}\left[Ra\, a^2-(a^2+n^2\pi^2)^2\right].
$$
Then:
\begin{itemize}
\item[1.] when $\mathcal C_2<\mathcal C_1$, i.e. 
$Ra<\dis (a^2+n^2\pi^2)^2/a^2$,
 one falls into the regime of stability for the  Darcy-B\'enard problem with standard diffusion;
 \item[2.] when  $\mathcal C_2>\mathcal C_1$, i.e. 
$Ra>\dis (a^2+n^2\pi^2)^2/a^2$, standard diffusion in the Darcy-B\'enard problem leads to instability, while a generic superdiffusion implies stability and the existence of a transient growth of perturbations before they revert to zero.
\end{itemize}
Analytical results show that the greater the anomalous diffusion (i.e., the faster the asymptotic behavior of the function $h(t)$), the smaller the peak observed in the growth of perturbations during the transient regime.
\\
The main conclusions of the present study are listed below:

\begin{itemize}
    \item Being $h(t)$ the memory function of the diffusion process, if $\lim_{t\to\infty} \dis H(t)/t=\infty$, an asymptotic stability is observed both from a linear and a nonlinear point of view, with $H(t)$ being the primitive of $h(t)$.
    \item For all memory functions tested here an asymptotic stability was observed.
    \item Disturbances always present a transient growth for the power-law memory function. Such a transient growth is stronger and longer for $\mathcal{C}_2>\mathcal{C}_1$. Increasing the parameter $r$ decreases the transient growth magnitude and reduces the transient growth interval.
    \item Disturbances can present a transient growth behaviour for the exponential memory function case just for $\mathcal{C}_2>\mathcal{C}_1$. The case $\mathcal{C}_1>\mathcal{C}_2$ leads always to a monotonic temporal behaviour. 
    \item Disturbances present a transient growth for the case of logarithmic memory function when $\mathcal{C}_2>\mathcal{C}_1 \log(a)$ in the regime tested here. Such a transient growth is stronger and longer for $\mathcal{C}_2>\mathcal{C}_1$ as in the case of power-law memory function. If the transient growth is observed, an increasing $a$ means a decreasing maximum growth and a decreasing time necessary to reach the maximum.
\end{itemize}

\section*{Acknowledgments} 
This paper has been performed under the auspices of the GNFM of INdAM.
R. De Luca and F. Capone acknowledge the support of National Recovery and Resilience Plan (NRRP) funded by the European Union - NextGenerationEU - Project Title ``Mathematical Modeling of Biodiversity in the Mediterranean sea: from bacteria to predators, from meadows to currents" - project code $P202254HT8$ (CUP $B53D23027760001$) and the support of grant no. MUR-PRIN PNNR $2022$ - Project Title ``Modelling complex biOlogical systeMs for biofuEl productioN and sTorAge: mathematics meets green industry" - project code $202248TY47$ (CUP $E53D23005430006$). P. V. Brand\~ao acknowledges the support of Horizon Europe CoBRAIN (GA No. 101092211).




\end{document}